\documentclass{article}

\usepackage{PRIMEarxiv}

\usepackage[utf8]{inputenc} % allow utf-8 input
\usepackage[T1]{fontenc}    % use 8-bit T1 fonts
\usepackage{hyperref}       % hyperlinks
\usepackage{url}            % simple URL typesetting
\usepackage{booktabs}       % professional-quality tables
\usepackage{amsfonts}       % blackboard math symbols
\usepackage{nicefrac}       % compact symbols for 1/2, etc.
\usepackage{microtype}      % microtypography
\usepackage{lipsum}
\usepackage{fancyhdr}       % header
\usepackage{graphicx}       % graphics
\graphicspath{{media/}}     % organize your images and other 
\usepackage{multirow}
\usepackage{biblatex}
\title{Tissue Concepts: supervised foundation models in computational pathology
}

\author{
  Till Nicke\\
  Fraunhofer Institute for Digital Medicine MEVIS \\
  Bremen / Lübeck / Aachen, Germany\\
  \texttt{till.nicke@mevis.fraunhofer.de} \\
  \And
  Jan Raphael Schäfer\\
  Fraunhofer Institute for Digital Medicine MEVIS\\
  Bremen / Lübeck / Aachen, Germany \\
  \And
   Henning Höfener \\
  Fraunhofer Institute for Digital Medicine MEVIS\\
  Bremen / Lübeck / Aachen, Germany \\
  \AND
  Friedrich Feuerhake \\
  Institute for Pathology,\\
  Hannover Medical School, Hannover, Germany\\
  \And
  Dorit Merhof \\
  Institute of Image Analysis and Computer Vision,\\
  University of Regensburg, Regensburg, Germany
  \And
  Fabian Kießling \\
  Institute for Experimental Molecular Imaging,\\
  RWTH Aachen University, Aachen, Germany
  \And
  Johannes Lotz \\
  Fraunhofer Institute for Digital Medicine MEVIS \\
  Bremen / Lübeck / Aachen, Germany\\
}

\begin{document}
\maketitle

\begin{abstract}
Due to the increasing workload of pathologists, the need for automation to support diagnostic tasks and quantitative biomarker evaluation is becoming more and more apparent.
Foundation models have the potential to improve generalizability within and across centers and serve as starting points for data efficient development of specialized yet robust AI models.
However, the training foundation models themselves is usually very expensive in terms of data, computation, and time.

This paper proposes a supervised training method that drastically reduces these expenses.
The proposed method is based on multi-task learning to train a joint encoder, by combining 16 different classification, segmentation, and detection tasks on a total of 912,000 patches.
Since the encoder is capable of capturing the properties of the samples, we term it the \textit{Tissue Concepts} encoder.

To evaluate the performance and generalizability of the Tissue Concepts encoder across centers, classification of whole slide images from four of the most prevalent solid cancers - breast, colon, lung, and prostate - was used.
The experiments show that the Tissue Concepts model achieve comparable performance to models trained with self-supervision, while requiring only 6\% of the amount of training patches.
Furthermore, the Tissue Concepts encoder outperforms an ImageNet pre-trained encoder on both in-domain and out-of-domain data.

The pre-trained models and will be made available under \url{https://github.com/FraunhoferMEVIS/MedicalMultitaskModeling}.
\end{abstract}

% keywords can be removed

\section{Introduction}
\label{sec:into}
The need for diagnostic systems to help pathologists manage the anticipated workload increases as cancer cancers worldwide are on the rise \cite{moscalu_histopathological_2023}. As \cite{sung_global_2021} estimate, breast, colorectal, prostate, and lung cancers are among the six most common cancers types. Projections suggest that cases of these cancers will continue to increase, posing significant challenges due to time-consuming diagnosis, increased demand for tumor subtyping, and personalized treatment \cite{rahib_estimated_2021, soerjomataram_planning_2021, siegel_cancer_2023}. Deep learning (DL) has made significant progress in medical imaging, particularly in the field of computational pathology (CPath). Some studies have demonstrated that DL models even surpass human performance in certain tasks, making them effective tools to help pathologists cope with the increasing workload \cite{zhang_pathologist-level_2019, liu_deep_2020}. However, the unavailability of the required large data sets and the needed investment of time and effort limits the effectiveness and impact of DL models in pathology.

Recent advances in self-supervised learning have enabled the training of deep neural networks on large amounts of unlabeled medical data, resulting in the creation of foundation models in computer vision \cite{zhang_challenges_2023}. These models are pre-trained on a wide range of images, primarily using self-supervision through contrastive learning or masked image modeling. They have been shown to perform well in downstream tasks, including patch classification, and weakly labeled whole slide image (WSI) classification \cite{wang_transformer-based_2022, campanella_computational_2023}. Projects such as the Tissue Cancer Genome Atlas Program (TCGA) provide an openly available data source of thousands of WSIs for training these networks on real-world data. This vast amount of data is necessary for self-supervised trained networks to reach their full potential \cite{zhang_challenges_2023}. However, the amount of resources required to create, train, and deploy such models has raised concerns among researchers about environmental and other impacts \cite{dhar_carbon_2020, kaack_aligning_2022, the_lancet_digital_health_curbing_2023}. In addition, extended training periods of several weeks impede development cycles and prolong research time.

Supervised learning, on the other hand, has been shown to outperform models trained on self-supervision in some tasks \cite{pototzky_does_2022, huang_segment_2024}. Although there are many annotated datasets available through challenges or other benchmarks, these datasets vary in size and contain annotations with varying degrees of detail. This variability between the datasets makes it challenging to condense the knowledge they contain into a single model. One approach to integrating all of these label types is to use multi-task learning (MTL) \cite{caruana_multitask_1997}. In \cite{schafer_overcoming_2023}, we recently proposed a learning framework that combines the information contained in different labeling strategies, including detection, segmentation, and classification, and use it to train a single shared backbone model on a large corpus of images. In the study, images from different medical imaging domains, such as CT, X-ray, and microscopic images but also non-medical images were included.

\begin{figure*}[t]
    \centering
    \includegraphics[width=\textwidth]{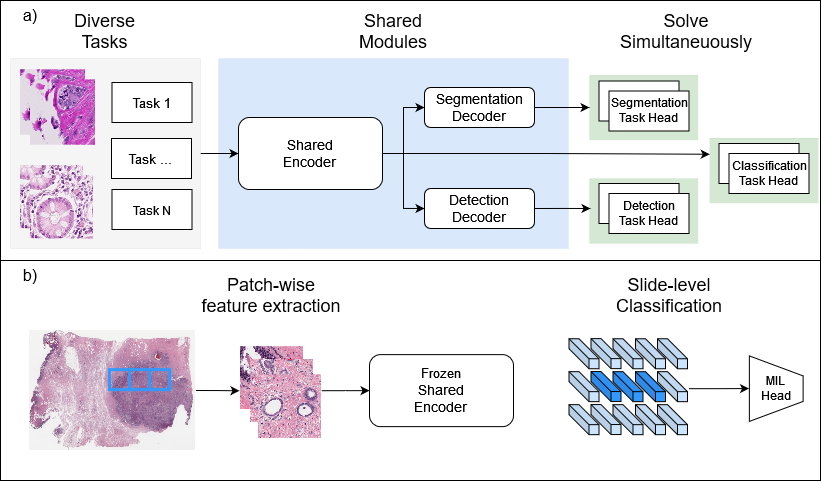}
    \caption{Overview of the study. a) Different pre-training of Tissue Concepts using multi-task learning on 16 different tasks. b) the shared encoder is evaluated using multiple-instance learning on WSI classification. From each WSI, patches of size $224 \times 224$ are extracted in an iterative windowing fashion and the latent representation is positioned at the same spatial location as the patches. A simple CNN is trained on the latent WSIs to learn the label at the slice level.\\
    }
    \label{fig:overview}
\end{figure*}
This paper demonstrates that training a foundation model on supervised signals in CPath using MTL requires less data, time, and energy compared to models trained with self-supervision. At the same time, the measured performance is similar to that obtained from models trained on about 17 times more data without supervision. Following the MTL training scheme presented in \autoref{fig:overview}a, this paper presents \textit{Tissue Concepts} (TC), a robust encoder that is trained on a mixture of diverse annotations from small and medium-sized datasets in CPath to learn different concepts related to tissue. Considering the future prediction of cancer cases and clinical workflow, we evaluated the performance of the encoder on the four major cancer types, breast, colon, lung, and prostate, for classification of entire slide images, as shown in \autoref{fig:overview}b. In addition, since models trained on one site are known to perform worse when evaluated on different sites, we tested the performance of Tissue Concepts using a cross-center evaluation scheme \cite{howard_impact_2021}.

The main contributions of the paper can be summarized as follows.
\begin{itemize}
    \item We show that diverse pre-training using MTL learns robust representations and drastically reduces the required amount of data compared to self-supervised approaches.
    \item Our evaluation of the Tissue Concepts encoder on four of the most prevalent cancer types across multiple centers highlights the generalizability of our approach.
\end{itemize}

\section{Related Work}
\label{sec:related_work}
First approaches using MTL in CPath were presented by\cite{mormont_multi-task_2021} and \cite{graham_one_2023}. Mormont and colleagues converted different datasets into 22 classification tasks to train a shared network and contrasted the learned encoders against ImageNet weights. The latent representations of the encoder were used to train an SVM. They found that the representations perform equally or better than the baseline ImageNet weights. Graham et al. then used MTL on segmentation and classification tasks. This research focused on specific tasks that were present in the pre-training. 
However, the evaluation of the general-purpose encoder and the corresponding latent representations based on whole slide image classification combined with cross-center evaluation is still an unexplored area. In addition, general purpose encoders in the form of foundation models have not been considered by \cite{graham_one_2023}.

In \cite{schafer_overcoming_2023}, we presented a first approach using MTL to train supervised foundation models. Using expert knowledge in the form of multi-task learning we trained a shared model, called UMedPT, which can be applied to various medical images. To achieve this, different imaging domains, such as CT, X-ray, and microscopic images, were used to train a shared backbone on classification, segmentation, and detection tasks. Currently, the impact of tasks outside the histopathology domain remains unclear due to the diverse pre-training of the encoder. This impact on performance and robustness requires further investigation.

The following sections focus in more detail on two topics discussed in this paper. Although foundation models are still largely unexplored in terms of their application and performance, some approaches are mentioned below.

\subsection{Foundation Models}
A foundation model is broadly defined as being trained on a wide variety of data and being easily adaptable to many different downstream tasks \cite{bommasani_opportunities_2022, fei_towards_2022}. 

\cite{wang_transformer-based_2022} used data from the TCGA in combination with data from the pathology AI platform, PAIP, to train a modified swin transformer, called CTransPath (CTP), using self-supervision on 15 million patches. They presented an adapted contrastive loss, based on MoCo v3 \cite{chen_empirical_2021}, which uses a memory bank to retrieve the top S semantically relevant entries. These entries were used as additional positive examples for the loss calculation. The authors evaluated their model using patch classification, image retrieval, and weakly labeled WSI classification. Due to the large number of training images and the slow convergence of self-supervised training, they reported a training time of 250 hours on 48 GPUs (12.000 GPU-hours). 

\cite{campanella_computational_2023}, presented a comparable approach, training a tiny vision transformer (ViT) using standard DINO, and a ViT base model using standard masked autoencoding (MAE), both trained on about 3 billion patches. The authors report a training time of over 3000 GPU-hours for the models that were evaluated on a variety of tasks ranging from disease detection to outcome prediction. The evaluation also included images scanned at a different hospital from the training slides. 

\cite{chen_towards_2024} present a general purpose foundation model that leverages over 100 million patches from 100.000 WSI slides across 20 major cancer types. They train a large ViT on patches, that were collected in an internal dataset, using DINOv2 \cite{oquab2023dinov2}. They evaluate the model on 34 Tasks and find that it surpasses the previous baselines on most of them. The model was trained on 24 80GB GPUs.

Overall, all of the presented models rely on large image databases and require long training times, which contributes to increased CO\textsubscript{2} emissions. The presented TC encoder and MTL training aim to reduce the need for large amounts of data while maintaining the desired performance.
In addition, cross-center evaluation is needed to accurately asses models' performances. 

\subsection{Weakly Labeled WSI Classification}
\label{sec:wsi_clf}
Learning from WSIs that are only labeled on a case basis, or that have only one endpoint, is challenging because training on the entire image at once typically exceeds the GPU memory. In addition, since a WSI provides only one sample, many WSIs are needed to effectively train a deep learning model. Classification of such gigapixel images is therefore typically performed using multiple instance learning (MIL) \cite{campanella_clinical-grade_2019, ghaffari_laleh_benchmarking_2022}. Using MIL involves two parts: first, extracting features from patches of the WSI using a pre-trained encoder to convert them into their latent representations, and second, aggregating features from a WSI using a trainable MIL head to predict the given label \cite{liu_advmil_2024}. Therefore, robust encoders are needed to obtain patch representations that facilitate the second step of MIL \cite{godson_immune_2024}. In this paper, MIL is used as an evaluation procedure to test the representativeness of the encoder's features. The following presents the most commonly used approaches that focus on solving the second stage of MIL using either attention or convolution-based methods.

\cite{lu_data-efficient_2021} introduced CLAM, a clustering-constrained attention MIL algorithm. The authors trained an attention-based head on features extracted from patches of a WSI to classify the corresponding labels. The attention was then used to identify subregions of high diagnostic value, which in turn were used to classify the entire slide. In addition, instance-level clustering was applied over the representative regions to constrain and refine the feature space.

\cite{shao_transmil_2021} proposed TransMIL, an attention-based correlation method for solving weakly labeled classification tasks. The method uses differently sized convolutional layers to apply additional pyramid position encoding information between the attention modules. This allows the attention layers to aggregate morphological features, while the Pyramid Position Encoding Generator (PPEG) encodes spatial information.

\cite{tellez_neural_2021} proposed neural image compression to train on entire WSIs. The authors trained an autoencoder on patches and used the resulting encoder for feature extraction. The patches extracted in the image domain were encoded and their latent representations were placed in the same spatial location. This effectively compressed the entire WSI into a smaller latent image with more channels, while preserving the spatial relationship between the individual patches. A small CNN was then trained on the compressed WSIs to predict the label of the WSI. In a second version of this approach the same authors used multi-task learning on four classification tasks to train the feature extractor \cite{tellez_extending_2020}. The effect of segmentation and detection tasks, as well as more diverse pre-training, remained a point of further investigation and are part of the research presented in this paper. 

All presented methods propose different aggregation methods to learn the desired label predictions based on the extracted features and thus work with the features extracted by the TC encoder. As an evaluation method, we adapted the convolution-based aggregation method presented by \cite{tellez_neural_2021} and also applied an attention-based approach based on \cite{ilse2018attention}, which are further described in Section \ref{subsec:mil}. 

\section{Methods}
\label{sec:method}
Multi-task learning was used to train the Tissue Concepts encoder on 14 small and medium-sized datasets. The pre-training datasets and the procedure used during this phase are described in the following sections. In addition, the evaluation datasets, MIL head, and corresponding training are explained in detail.

\subsection{Pre-training Datasets}
\label{sec:DS}
\begin{table*}
\centering
\caption{\label{tab:datasets} Overview of the datasets used in the training of the “Tissue Concepts” encoder (purpose “training”), the ablation study (“ablation”), and in the validation experiments (“validation”).}
\begin{tabular}{|l|c|c|c|c|c|}
\hline
Dataset &  \# WSIs & \# Patches & Organ & Task Type &  Purpose\\
\hline
NCT-CRC-HE 100K \cite{kather_predicting_2019} & 86 & 100000 & Colorectal & Classification &  training\\
Crag \cite{graham_mild-net_2019, awan_glandular_2017}& 14 & 173  & Colorectal & Segmentation &  training\\
SemiCOL \cite{semicolHomeSemiCOL} & 20 & 1759 & Colorectal & Segmentation&  training\\
Conic \cite{graham_conic_2024, graham_lizard_2021}& 291 & 4981  & Colon & Detection &  training\\
Conic \cite{graham_conic_2024, graham_lizard_2021}& 291 & 4981  & Colon & Segmentation &  training\\
Arvaniti \cite{arvaniti_automated_2018}& 641 &  2000 & Prostate & Classification &  training\\
Peso \cite{bulten_epithelium_2019} & 62 & 392 & Prostate & Segmentation &  training\\
Schoemig-Markiefka \cite{schomig-markiefka_quality_2021} & 1177 & 600000 & Prostate & Classification&  training\\
PANDA \cite{bulten_artificial_2022}& 2000 & 40000  & Prostate & Classification &  training\\
PANDA \cite{bulten_artificial_2022}& 2000 & 40000  & Prostate & Segmentation &  training\\
TUH \cite{wevodau_temple_2021}& 136 & 3720  & Breast & Classification &  training\\
TIGER \cite{shephard_tiager_2022} & 151 & 1876 & Breast & Segmentation &  training\\

BCSS \cite{amgad_structured_2019}& 18 & 139  & Breast & Segmentation &  training\\
BreakHis \cite{spanhol_dataset_2016}& 82 & 4076  & Breast & Classification &  training\\
MiDoG \cite{aubreville_mitosis_2023}& 49 & 405  & Various & Detection &  training\\
HubMap \cite{jain_segmenting_2023} & 315 & 1260 & Various & Segmentation &  training\\
\hline
\textbf{Sum} & 7042 & 912157 & Various & Various & training\\
\hline
Bach \cite{aresta_bach_2019} & 400 & 400 & Breast & Classification & ablation \\
\hline
BRACS \cite{brancati_bracs_2022} & 547 & 5566906 & Breast & MIL & validation\\
Panda \cite{bulten_artificial_2022} & 10000 & 98928234 & Prostate & MIL & validation\\
SemiCOL \cite{semicolHomeSemiCOL} & 499 & 1436065 & Colorectal & MIL & validation\\
TCGA-NSCLC \cite{tcga-lusc, tcga-luad} & 1006 & 12641437 & Lung & MIL & validation\\
\hline
Sum & 12052 & 118.572.642 & Various & MIL & validation\\
\hline
\end{tabular}
\end{table*}
To pre-train the shared encoder, a total of 14 data sources were collected and distributed over 16 tasks. All data sources with corresponding number of patches and WSIs, as well as their tasks are presented in \autoref{tab:datasets}. The NCT-CRC-HE 100k, Panda, TUH, Breakhis, and Arvaniti datasets have been designated as patch-level classification tasks by their respective curators. Data from the PANDA dataset was used for classification and segmentation tasks, where the patches in PANDA were 20 patches per WSI, sampled of a total of 4000 WSIs. To increase data diversity, the Breakhis dataset was used at both 40x and 100x magnification.
Conic and MiDoG were included as detection tasks, while Conic also served as a segmentation task. Other datasets used for segmentation were the SemiCOL training dataset, Arvaniti, Peso, Schoemig-Markiefka, PANDA, TIGER, CRAG, BCSS, and HubMap. Each of the mentioned datasets is described in more detail in the \autoref{app:DS}. In total, about 912,000 patches from around 7,000 WSIs were used during the training. The training data consisted of about 100,000 patches from colorectal, 600,000 patches from prostate, and 10,000 patches from breast tissue. In addition, a small number of spleen, liver, and skin tissue slides were included, which are marked as "various" in \autoref{tab:datasets}.

All patches were extracted or scaled to $224 \times 224$ pixels, resulting in a resolution of approximately 1 to 0.5 micron per pixel (MPP). In addition, standard augmentation, such as random rotation, distortion, blurring, brightness, contrast, and hue changes were applied during pre-training.

\subsection{Encoder Training}

The shared backbone was trained using multi-task learning with all of the above data sources \cite{schafer_overcoming_2023}. In the MTL pipeline, each task was solved by a shallow task-specific head (\(\theta_{t}\)) that received input from larger, shared modules (\(\theta_{shared}\)). \autoref{fig:overview}a gives a brief overview of the different blocks in the architecture. The shared blocks are shown in the blue area, while the individual tasks are shown in green.
During training, all tasks were treated equally and were processed iteratively within the training loop. For each task the task-specific loss was computed and accumulated on the total loss. Formally, the total loss \(\mathcal{L}_{total}(X, Y)\) for a set of tasks \(X=(X_1,...,X_T)\) of task specific images  \(X_t=(x_{t,1},...,x_{t,N})\) and corresponding labels \(Y_{t}=(y_{t,1},...,y_{t,N})\) is computed as
\begin{equation}
\label{eq:mtl_loss}
\mathcal{L}_{total}(X, Y) = \sum_{t=1}^{T} \mathcal{L}_{t}(\theta_{t}(\theta_{shared}(X_{t})), Y_{t}) 
\end{equation}
where the loss of the task \(t\) is calculated based on the output of the task specific head $\theta_{t}$ that received its' input from a shared network structure \(\theta_{shared}\). \(X_{t}\) and \(Y_{t}\) describe the task-specific batch input images and the corresponding labels respectively. While the task-specific parameters \(\theta_{t}\) in \autoref{eq:mtl_loss} changed, depending on which task was being processed, the shared parameters \(\theta_{shared}\) were task-independent. Since all tasks contributed equally to the total loss, multiple cycles through all tasks can be performed by performing gradient accumulation before an optimization step is performed.

The tasks themselves were divided into classification, segmentation, and detection. All tasks shared the same encoder, which yielded a feature pyramid with four feature maps at different perceptual levels. The feature pyramid was further processed depending on the task. 

\textbf{Classification tasks} received a globally average pooled version of the lowest feature pyramid level from the encoder. Each task-specific classification head consisted of 20\% dropout followed by a linear layer, mapping from the dimension of the latent space to the number of classes. By minimizing the cross-entropy loss, each head learned to predict the task-specific classes. 

\textbf{Segmentation tasks} consisted of a shared u-net-style decoder that received features from all levels of the shared encoder. Each individual segmentation task was then processed by a \(1 \times 1\) convolutional layer operating on the output of the last decoder level. A combination of dice and focal loss was minimized by each segmentation head. 

The \textbf{detection tasks} also consisted of a shared u-net style decoder that received features from all levels of the shared encoder to extract spatially relevant feature maps. Unlike the segmentation tasks head, the detection task head consisted of an FCOS detection head, which minimized the corresponding losses.  

Since transformer-based approaches are currently dominant in the literature we chose a tiny swin transformer as shared encoder \cite{liu_swin_2021, azad_advances_2024} (further denoted as TC-Swin). The encoder shares the same latent dimension with CTransPath, 768, and has 27.5M trainable parameters. Additionally, we compared the attention-based backbone to a tiny ConvNeXt architecture that shares a similar number of parameters and latent dimension but follows a convolution-based approach \cite{liu_convnet_2022} (further denoted as TC-Conv). Both networks were initialized with the corresponding ImageNet-1k weights which were imported from the torchvision library \cite{russakovsky_imagenet_2015}. During pre-training, the networks were optimized for 5 million steps until convergence, where one step was defined as computing the loss for one batch of one task. AdamW optimizer \cite{loshchilov_decoupled_2019} was used with a constant learning rate of $10^{-4}$, a weight decay of 0.01, and a gradient accumulation of 128 steps. In total, the models were trained for 160 hours on one NVidia RTX A5000 GPU and the final models were used for evaluation.

\section{Evaluation}
This section first establishes a comparison between UMedPT and TC using a sample efficiency experiment as an ablation study. Then, the evaluation datasets and corresponding tasks are presented, and the MIL head training procedure is described.

\subsection{Ablation Study}
\label{sec:ablation}
To determine whether a multi-domain encoder, UMedPT, or a histology specific encoder, TC-Swin, performs better on unseen histology data, we measured the sample efficiency on an unseen, in-domain dataset. In this experiment, the frozen encoder generated latent representations for all patches in the downstream dataset. These latent representations were then used as input to a random forest model. Sample efficiency was measured by establishing of a fixed training and validation split. Different subsets of images per class were systematically sampled from the training split to serve as training examples for the random forest. Each model was consecutively evaluated on the designated test split. This training-validation process was repeated 10 times with different seed values for robustness in the analyses and the F1-score was reported. We compared TC-Swin to the base Swin transformer encoder presented in \cite{schafer_overcoming_2023}. Additionally, ImageNet weights from torchvision were used as baseline for a tiny swin transformer.

The \textbf{BACH} challenge dataset \cite{aresta_bach_2019} is a breast tumor classification dataset consisting of 400 images of \(2048 \times 1536\) pixels at 0.25 MPP, which are equally divided into 4 classes. Images were center cropped to \(1024 \times 1024\) pixels and downsampled to \(224 \times 224\) pixels. The dataset was split 80/20 into a training and test subset. From the training split of the BACH dataset, sets of 1, 3, 5, 10, 25, and \textit{max} 46 images per class were sampled using 10 different seeds. For each training, the random forest was evaluated on the left out test set.
\begin{figure}[t]
    \centering
    \includegraphics[width=.65\linewidth]{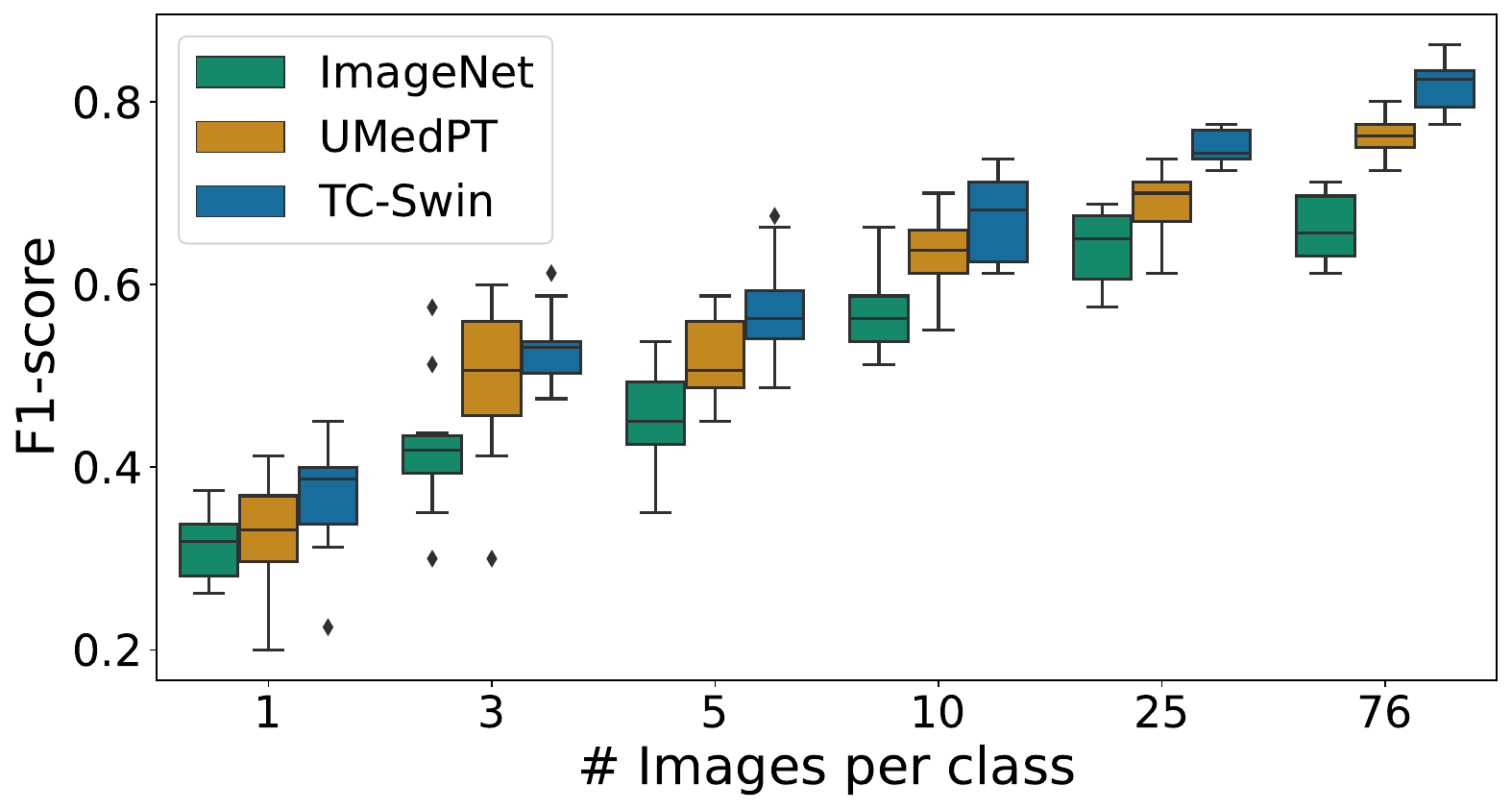}
    \caption{Sample efficiency for encoders of different specificity on the BACH patch classification dataset. Each boxplot represents 10 repetitions. The TC-Swin encoder is compared to the tiny swin transformer ImageNet weights and a small swin transformer of UMedPT \cite{schafer_overcoming_2023}. The F1-scores are plotted against increasing numbers of images per class to examine sample efficiency.\\
    }
    \label{fig:ablation}
\end{figure}

Figure \ref{fig:ablation} shows that the domain-specific encoder outperforms the multi-domain encoder and the pre-trained ImageNet encoder in terms of sample efficiency.
The TC-Swin encoder showed higher F1-scores compared to the multi-domain encoder and to ImageNet. With three images per class, TC-Swin's mean F1-score was at 0.531, while UMedPT and ImageNet were at 0.496 and 0.423, respectively. With 10 images per class, the TC-Swin encoder showed a mean F1-score of 0.673, while UmedPT and ImageNet showed values of 0.628 and 0.566, respectively. Using the maximum number of available training images per class, the differences in reported performance are similar. While TC-Swin achieved 0.816, UmedPT and ImageNet achieved mean F1-scores of 0.762 and 0.662, respectively.

The TC-Swin encoder outperforms the multi-domain medical encoder UMedPT even though the later was trained with more images overall. In particular, with 5 or more images per class, the features obtained from the TC encoder provide better performance than the ImageNet and UMedPT features. When considering 1 and 3 images per class, UMedPT shows similar performance to the TC encoder, but also shows a higher variance over the 10 seeds. Therefore, we excluded the UMedPT encoder from further evaluation and focused on the TC encoder instead.

\subsection{Evaluation datasets}
We compared the TC encoder to the current state of the art for each organ, referred to as the external baseline, as well as to the results obtained using the features of the publicly available CTransPath encoder \cite{wang_transformer-based_2022}. In addition, we use the Swin transformer model with ImageNet weights without further training as an additional baseline.

The learned representations of Tissue Concepts were evaluated using weakly labeled WSI classification tasks. Breast, prostate, and caolorectal tissue represent tissues within the training domain, lung cancer tissue was deliberately chosen to represent tissues outside the training domain. The architecture and learning schedule as described in Section \ref{subsec:mil}, was used during the training of each MIL head. Where possible, multi-site datasets were split so that evaluation sites were not included in the training. The feature extractors were evaluated on the following datasets for each organ: 

\textbf{Breast Cancer}: The BReAst Cancer Subtyping (BRACS) dataset consists of 547 WSIs from 189 patients which are already divided into training (395), validation (87) and test (65) splits \cite{brancati_bracs_2022}. According to \cite{pati_hierarchical_2022}, the WSIs were obtained at the Department of Pathology at the National Cancer Institute - IRCCS-Fondazione Pascale, Naples, Italy, and were scanned with an Aperio AT2 scanner at a resolution of 0.25 microns per pixel (mpp). Each WSI is assigned to one of 7 different classes: normal, benign, usual ductal hyperplasia (UDH), atypical ductal hyperplasia (ADH), flat epithelial atypia (FEA), ductal carcinoma in situ (DCIS), and invasive. In addition, \cite{pati_hierarchical_2022} grouped the different types of breast carcinoma into four coarser classes: normal, non-cancerous, precancerous, and cancerous. Furthermore, binary classification between invasive and non-invasive breast cancer was also performed by \cite{pati_hierarchical_2022}. The authors also presented a trained model, HACT-Net, and applied it to this dataset. They reported the weighted F1-score for all of the classification problems mentioned above which we used as an external baseline and also reported the weighted F1-score in the BRACS experiments. Additionally, following \cite{maier-hein_metrics_2024}, we reported accuracy as multi-class metric and area under the receiver operating curve (AUC) as multi-threshold metric. For the evaluation of the TC-encoder, three experiments were created form the dataset, similarly to the approach proposed by \cite{pati_hierarchical_2022}. The first experiment consisted of differentiating between normal and cancerous tissue slides. The second experiment focused on more fine-grained classification between normal, non-cancerous, precancerous, and cancerous tissues. Finally, the third experiment for breast cancer aimed at predicting one of the seven sub-types provided.

\textbf{Prostate Cancer}: The Prostate cANcer graDe Assessment (PANDA) challenge provides 10616 openly available WSIs from two different centers, Karolinska and Radboud \cite{bulten_artificial_2022}. Each of the WSIs was assigned an ISUP score resulting from the primary and secondary Gleason pattern. The score ranges from 0 to 5, with 0 representing normal tissue and higher numbers representing more severe prostate cancer. This information was used to create binary classification task or evaluation, where the task consists of differentiating between normal (ISUP 0) and cancerous tissue (ISUP $>$ 0). 
To test cross-center performance and overall performance, three different experiments were created from this dataset. First, the entire set of images obtained from the Radboud UMC was used as both the training and validation set, using a split ratio of 90\% for training and 10\% for validation. All images provided by the Karolinska Institute were used as the hold-out test set. In a subsequent experiment, the roles of training and test centers were swapped. In addition, a 5-fold cross-validation experiment involving both centers was conducted. We note that a small percentage of the dataset in the form of pre-selected patches were used during pre-training of the encoder, however, the task-specific head was discarded. To evaluate the effect of possible data leakage \cite{kapoor_leakage_2023} an additional encoder was trained without the PANDA dataset in pre-training and evaluated only on this task. \cite{mahdi_behzadi_weakly-supervised_2022} used this dataset in a k-fold cross-validation to evaluate the performance of their model and reported the accuracy and area under the curve for binary classification. Their reported results serve as an external baseline. \cite{mahdi_behzadi_weakly-supervised_2022} reported the accuracy and AUC of their algorithm on the dataset. To be comparable, we also report these metrics in our experiments. Additionally, we report the F1-score as per-class metric, as suggested by \cite{maier-hein_metrics_2024}.

\textbf{Colorectal Cancer}: During the SemiCOL challenge, a dataset of 499 WSIs from 4 different centers was provided as training dataset. The challenge task was to predict whether a given slide contained cancerous tissue or normal tissue. During the challenge, we used a modified model presented in \cite{schafer_overcoming_2023} as an initial starting point and further fine-tuned on several colon and colorectal tasks. Using the learned encoder to extract features from the WSI, a MIL head was successively trained on these features. This resulted in an external test AUC of 0.99. In the experiments presented in this paper, the challenge training data was split by slide provider, effectively creating four smaller subsets of the original challenge dataset. For the evaluation presented in this paper, four different experiments were conducted. Images from one center were used as training examples (90/10), while images from the other three centers served as the hold out test set. By treating each center as training provider once, and using the other three centers as test centers, four different configurations were created. We note that a small percentage of patch data of the dataset was used during pre-training of the encoder, however, the task-specific head was discarded. To the best of our knowledge, there is no other publication reporting results on this specific dataset, so we are unable to report an external baseline. Since the dataset is balanced with regard to the class distribution for each center in all splits, we reported the Accuracy, F1-Score and AUC as metrics, following \cite{maier-hein_metrics_2024}.

\textbf{Lung Cancer}: The TCGA-NSCLC dataset contains two subtypes of lung cancer which are lung squamous cell carcinoma (TCGA-LUSC \cite{tcga-lusc}), and lung adenocarcinoma (TCGA-LUAD \cite{tcga-luad}) in 1006 (512 LUSC, 494 LUAD) slides from over 40 different sites. Two training and testing scenarios were created from the TCGA-NSCLC dataset. One experiment focused on corss-center evaluation. In this scenario, the three largest contributors (Johns Hopkins, International Genomics Consortium, and Asterand) were selected to serve as the training and validation subset providers. The remaining 39 sites were selected to serve as hold out test set. This split resulted in 340 training slides (90/10 split) and 666 test slides. In addition, a 5-fold cross-validation experiment was performed on the entire dataset to increase training set diversity, again using a 90/10 training and validation split. The objective of each experiment was to distinguish between slides containing lung adenocarcinoma and those containing squamous cell carcinoma (LUAD vs LUSC). The obtained results were compared to the results reported by \cite{shao_transmil_2021} and \cite{wang_transformer-based_2022}. To be comparable, the accuracy and AUC, as reported by the external baselines, were selected as metrics. Additionally, the F1-score was reported.

\subsection{Training of the MIL head}
\label{subsec:mil}
\begin{figure*}[t]
    \centering
    \includegraphics[width=.89\textwidth]{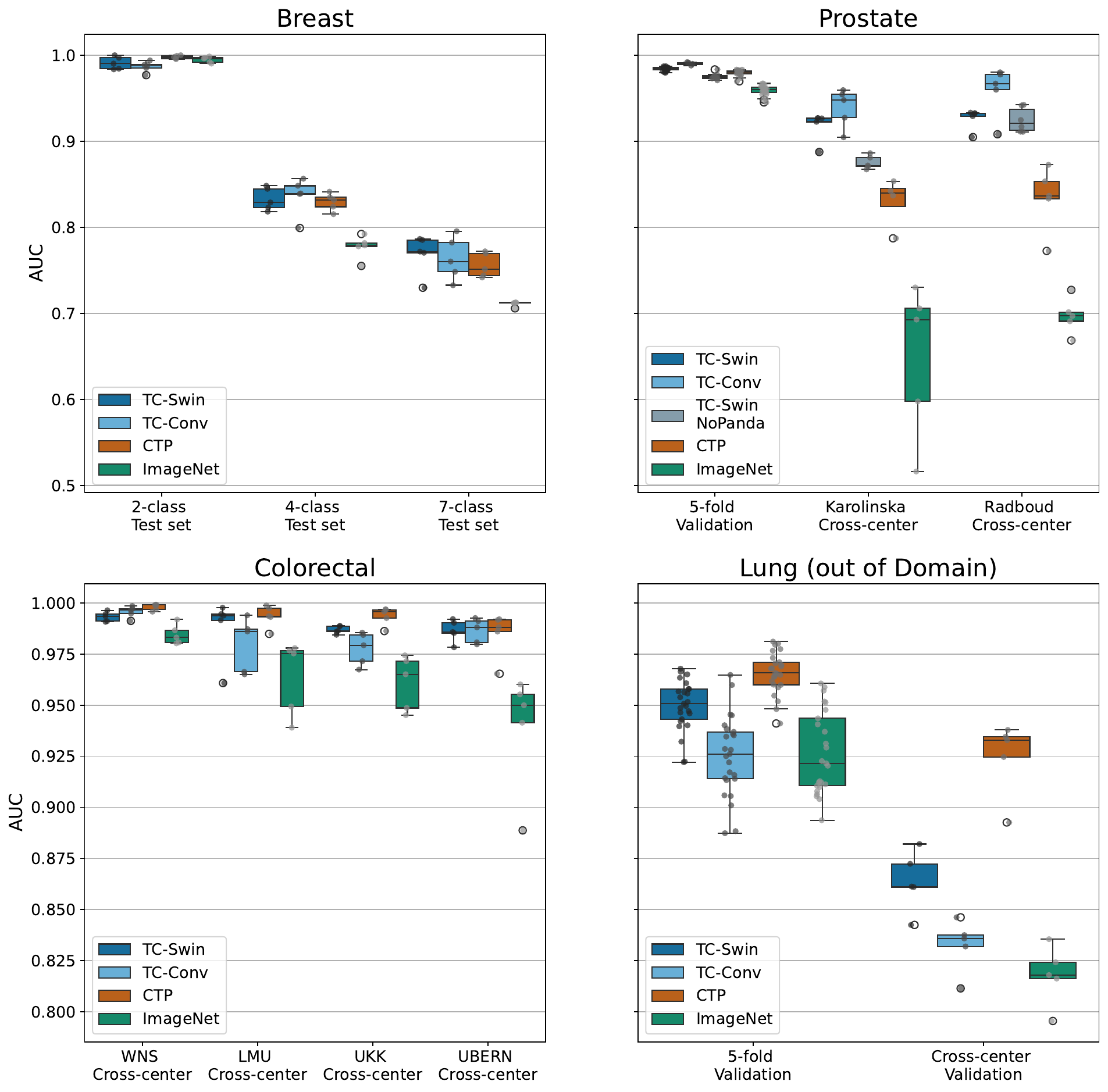}
    \caption{Organ-specific model performance when applied to downstream data. Models consist of organ-specific head and frozen encoders: ImageNet, CTP, TC-Swin, TC-Conv. While breast, colorectal, and prostate tissue was included in the pre-training, no lung tissue has been used during encoder pre-training of TC. "Breast" shows the results on three different classification tasks based on the BRACS dataset. "Colon" shows the cross-center performance between sites of the SemiCol challenge when trained on one center and evaluated on the others. "Lung" shows a 5-fold cross-validation and a cross-center evaluation. "Prostate" shows the performance on two cross-center evaluations as well as a 5-fold cross-validation from the Panda-challenge dataset. Boxes cover the inter-quartile-range, the median is marked by a horizontal line.\\
    }
    \label{fig:results}
\end{figure*}

Following medical diagnoses, WSIs are often only labeled at the patient or slide level. To address this, we used MIL to train slide-level classification heads for the respective evaluation tasks. The WSIs were compressed into latent images, by the respective frozen encoder, as suggested by \cite{tellez_neural_2021}. During this process, all patches of a WSI were extracted with a size of \(224 \times 224\) at about 0.5 MPP.

Using the exact same architecture for aggregation as presented by \cite{tellez_neural_2021}, we found that training the proposed MIL head left room for improvement in terms of training stability and convergence speed. The selection and adaptation were motivated by the prior performance on the SemiCOL challenge \cite{semicolHomeSemiCOL}. There, we fine-tuned UMedPT and trained the small MIL head on WSI classification. On an external, multi-center test set, this pipeline achieved an AUC of 0.998.

To aggregate the latent WSIs, a small CNN with global max pooling, was used. The initial layer utilized a 1x1 convolution, reducing the latent WSIs' 768 channel to 16. Following this, 3x3 convolutions with a stride of 2, and padding of 1 doubling in depth each time were applied. Instance normalization, leaky ReLU activation, and 10\% 2D dropout were applied in the convolutional layers. The lowest feature pyramid of 64 channels was globally max pooled, followed by one hidden linear layer and a classification layer. This resulted in 19250 total trainable parameters. The convolutional layers were initialized by drawing from a Kaiming normal distribution with a fixed seed. Additionally, we tested an attention-based approach by replacing the max pooling layer with an attention pooling layer, as suggested by \cite{ilse2018attention}. Overall, we found similar results with both approaches across all organs. The results using an ABMIL aggregation head can be found in the \autoref{sec:abmil}.

The same architecture was used over all evaluation datasets. To avoid overfitting, a label smoothing of 0.1 was used while minimizing the cross-entropy loss. The MIL head was trained using the AdamW optimizer with a constant learning rate of $10^{-4}$ and a weight decay of 0.01 for 100 epochs, while monitoring the validation loss. The best performing model was used for testing. During training, additional augmentations were applied to each latent WSI. Since each latent WSI was treated as a 768-channel image, random flipping, mirroring, and resizing with nearest neighbor interpolation were applied during training. This resulted in high-dimensional latent WSI representations with sizes ranging from \(32 \times 32\) to \(224 \times 224\).

\section{Results}
An overview of the obtained results is given in \autoref{fig:results}, where all tasks are presented with corresponding test AUC scores over 5 runs. The results show that the TC encoder, although trained with less data, performs as well as, or better than the CTransPath encoder, given the selected tasks. The Swin transformer with ImageNet weights exhibit the lowest overall performance and the largest variance across all tasks. All reported results were obtained using the default parameters of the corresponding functions from the scitkit-learn library \cite{pedregosa_scikit-learn_nodate}. In the following, the results on the different organs are examined in more detail.

\subsection{Breast Cancer}
\begin{table*}[t]
    \caption{Breast: Tissue-specific model performance when applied to the BRACS challenge data. The model consist of organ-specific head and frozen encoders: ImageNet, CTP, TC-Swin, TC-Conv. The table shows mean weighted F1 score, AUC and standard deviation over 5 different runs for different problem formulations. The best score is shown in bold.}
    \centering
    \begin{tabular}{c|c|c|c}
        Problem & Model &  wF1 & AUC \\
        \hline
         \multirow{5}{4em}{2-class} & \cite{pati_hierarchical_2022} & \textit{0.96} & - \\
         \cline{2-4}
         & ImageNet & $\mathbf{0.960} \pm 0.02$ & $0.994 \pm 0.003$  \\
         & CTP & $0.954 \pm 0.02$ & $\mathbf{0.997} \pm 0.001$ \\
         & TC-Swin & $0.951 \pm 0.02$ & $0.992 \pm 0.001$ \\
         & TC-Conv & $0.921 \pm 0.02$ & $0.989 \pm 0.006$\\
        \hline
        \multirow{5}{4em}{4-class} & \cite{pati_hierarchical_2022} & \textit{0.84} & - \\
         \cline{2-4}
         & ImageNet & $0.490 \pm 0.033$ & $0.777 \pm 0.01$  \\
         & CTP & $\mathbf{0.613} \pm 0.03$ & $0.829 \pm 0.01$ \\
         & TC-Swin & $0.545 \pm 0.04$ & $0.83 \pm 0.01$ \\
         & TC-Conv & $0.555 \pm 0.03$ & $\mathbf{0.837} \pm 0.02$\\
         \hline
         \multirow{5}{4em}{7-class} & \cite{pati_hierarchical_2022} & \textit{0.69} & - \\
         \cline{2-4}
         & ImageNet & $0.349 \pm 0.01$ & $0.711 \pm 0.002$  \\
         & CTP & $0.3989 \pm 0.02$& $0.755 \pm 0.01$ \\
         & TC-Swin & $\mathbf{0.405} \pm 0.02$ & $\mathbf{0.769} \pm 0.02$ \\
         & TC-Conv & $0.378 \pm 0.02$ & $0.764 \pm 0.02$\\
    \end{tabular}
    \label{tab:bracs}
\end{table*}
\autoref{tab:bracs} shows the results for the three sub-problems of the BRACS classification challenge, including the weighted F1-score and the AUC score for each. For the binary classification problem, ImageNet weights achieve the highest weighted F1-score on average with 0.960, which is comparable to the external baseline of 0.96. CTP and TC-Swin show similar performance with average wF1-scores of 0.954 and 0.951, respectively. The TC-Conv encoder performs almost three percentage points worse than the TC-Swin encoder. Overall, all tested encoders result in similar AUC values.

Considering the more fine-grained classification of the given WSIs, none of the encoders matches the reported performance of the external baseline HACT-Net with a weighted F1-score of 0.84. The two TC encoders outperform the ImageNet weights on the four- and seven-class problem formulations. On the four-class problem, CTP achieves a mean weighted F1-score of 0.613, approximately 6 percentage points higher than the two TC encoders. However, when considering the AUC, all of the foundation models perform equally with a mean AUC of about 0.83. For the more detailed seven-class problem, the TC-Swin encoder achieves the highest mean weighted F1 and AUC scores out of all the tested models with 0.40 and 0.76, respectively. 

Overall, the performance of all four encoders tested decreases significantly as the problem formulation becomes more fine-grained. Although the external baseline performance also declines, the decline is not as steep as that observed with the foundation models and ImageNet weights. 

\subsection{Prostate Cancer}
\begin{table*}[t]
    \caption{Prostate: Tissue-specific model performance when applied to the PANDA challenge data. The model consist of organ-specific head and frozen encoders: ImageNet, CTP, TC-Swin, TC-Conv, TC-Swin-NoPanda. The table shows mean weighted Accuracy, F1 score, AUC, and standard deviations over 5 different runs for 5-fold cross-validation and cross-center transfer performance. The best score is shown in bold.}
    \centering
    \begin{tabular}{c|c|c|c|c}
         Clinic & Model & Acc & F1 & AUC\\
         \hline
         external & \cite{mahdi_behzadi_weakly-supervised_2022}& 0.95 & 0.93 & -\\
         \hline
         \multirow{4}{4em}{5 fold} & ImgNet & 0.900 $\pm 0.01$ & 0.877 $\pm 0.01$ &  0.958 $\pm 0.01$\\
        & CTP & 0.936 $\pm 0.01$ & 0.921 $\pm 0.01$ & 0.979 $\pm 0.01$ \\
        & TC-Swin & $0.942 \pm 0.01$ & $0.928 \pm 0.01$ & $0.983 \pm 0.01$ \\
        & TC-Conv & $\mathbf{0.957} \pm 0.05$ & $\mathbf{0.947} \pm 0.06$ & $\mathbf{0.990} \pm 0.01$\\
        & TC-Swin-NoPanda & 0.932 $\pm$ 0.01 & 0.918 $\pm$ 0.01 & {0.975 $\pm$ 0.003} \\
        \hline
         \multirow{4}{4em}{Karolinska $\rightarrow$ Radboud} & ImgNet & 0.658 $\pm 0.15$ & 0.529 $\pm 0.08$ &  0.648 $\pm 0.08$\\
        & CTP & 0.819 $\pm 0.01$ & 0.524 $\pm 0.04$ & 0.825 $\pm 0.03$ \\
        & TC-Swin & 0.896 $\pm 0.01$ & $0.829 \pm 0.01$ & 0.914 $\pm 0.02$ \\
        & TC-Conv & $\mathbf{0.900} \pm 0.05$ & $\mathbf{0.845} \pm 0.06$ & $\mathbf{0.958} \pm 0.01$\\
        & TC-Swin-NoPanda & 0.865 $\pm$ 0.07 & 0.758 $\pm$ 0.08 & 0.924 $\pm$ 0.01 \\
        \hline
        \multirow{4}{4em}{Radboud $\rightarrow$ Karolinska} & ImgNet & 0.646 $\pm 0.01$ & 0.577 $\pm 0.04$ &  0.697 $\pm 0.02$\\
        & CTP & 0.678 $\pm 0.03$ & 0.512 $\pm 0.11$ & 0.834 $\pm 0.03$ \\
        & TC-Swin & $\mathbf{0.846} \pm 0.02$ & $\mathbf{0.842} \pm 0.02$ & 0.926 $\pm 0.01$ \\
        & TC-Conv & 0.764 $\pm 0.05$ & 0.646 $\pm 0.06$ & $\mathbf{0.958} \pm 0.01$\\
        & TC-Swin-NoPanda & 0.821 $\pm$ 0.07 & 0.807 $\pm$ 0.06 & 0.875 $\pm$ 0.007 \\
    \end{tabular}
    \label{tab:prostate}
\end{table*}
\autoref{tab:prostate} shows the results obtained of five runs on the PANDA dataset with cross-center and mixed evaluation. The 'clinic' column indicates different settings, where "Karolinska $\rightarrow$ Radboud" refers to the MIL head being trained on data from the Karolinska Institute and evaluated on data from the Radboud Clinic. The additional encoder TC-Swin-NoPanda was trained without PANDA data in the pretraining. 

When comparing the results obtained in the 5-fold cross-validation with the external baseline, both TC encoders show performances that were close to the external baseline, with TC-Conv slightly surpassing it with a mean accuracy of 0.957. The mean accuracies of CTP and ImageNet are 2 and 5 percentage points below the selected baseline, respectively. All foundation models outperform the pure ImageNet encoder by at least three percentage points. In terms of accuracy and AUC scores, two TC encoders slightly outperform the CTP weights by one percentage point with mean acc of 0.936 and mean auc of 0.979. When training the TC-Swin encoder without the PANDA dataset, a slight drop in performance can be observed, resulting in similar metrics compared to the CTP encoder.

In the cross-center evaluation, the performance of all models decreases compared to an in-center evaluation. CTP achieves a mean accuracy of 0.819 when transferring decision boundaries from the Karolinska site to Radboud subset. The opposite scenario results in a decrease of 14 percentage points (0.678 mean acc). Furthermore, an even steeper drop in mean accuracy and AUC from 5-fold cross-validation to cross-center validation is observed when using the ImageNet weights (from 0.900 to 0.658). All TC encoders are capable of extracting robust features. The TC-Swin encoder (from 0.942 to 0.846) shows more stable performance than the TC-Conv encoder (from 0.957 to 0.764) when compared to the metrics reported in the cross-center evaluation. The TC-Swin encoder trained without panda shows slightly lower performance on average to the TC-Swin encoder trained with PANDA. Independently of including the PANDA dataset in the pre-training, all TC-based encoders show a stronger cross-center performance over CTP.
 
\subsection{Colorectal Cancer}
\begin{table*}[t]
    \centering
    \caption{Colorectal: Tissue-specific model performance when applied to the SemiCOL challenge data. The model consist of organ-specific MIL head and frozen encoders: ImageNet, CTP, TC-Swin, TC-Conv. The table shows mean accuracy, F1 score, AUC, and the respective standard deviations over 5 different runs. Each model was trained on images from one site and evaluated on all images from the remaining sites.  The best score is shown in bold.}
    \begin{tabular}{c|c|c|c|c}
        Clinic & Model & Acc & F1 & Auc \\
        \hline
        \multirow{4}{4em}{LMU} & ImgNet & 0.877 $\pm 0.05$ & 0.875 $\pm 0.06$ &  0.961 $\pm 0.01$\\
        & CTP & 0.967 $\pm 0.01$ & 0.964 $\pm 0.01$ & 0.991 $\pm 0.004$ \\
        & TC-Swin & 0.948 $\pm 0.03$ & 0.948 $\pm 0.03$ & 0.987 $\pm 0.01$ \\
        & TC-Conv & 0.921 $\pm 0.02$ & 0.919 $\pm 0.02$ & 0.979 $\pm 0.01$\\
        \hline
        \multirow{4}{4em}{UBERN} & ImgNet & 0.734 $\pm 0.09$ & 0.709 $\pm 0.11$ & 0.939 $\pm 0.02$\\
        & CTP & 0.860 $\pm 0.08$ & 0.854 $\pm 0.09$ & 0.985 $\pm 0.01$ \\
        & TC-Swin & 0.943 $\pm 0.02$ & 0.944 $\pm 0.02$ & 0.987 $\pm 0.01$ \\
        & TC-Conv & 0.943 $\pm 0.01$ & 0.943 $\pm 0.01$ & 0.986 $\pm 0.01$\\
        \hline
        \multirow{4}{4em}{UKK} & ImgNet & 0.890 $\pm 0.02$ & 0.889 $\pm 0.02$ & 0.960 $\pm 0.01$\\
        & CTP & 0.962 $\pm 0.01$ & 0.962 $\pm 0.01$ & 0.993 $\pm 0.01$ \\
        & TC-Swin & 0.955 $\pm 0.01$ & 0.953 $\pm 0.01$ & 0.989 $\pm 0.01$ \\
        & TC-Conv & 0.925 $\pm 0.02$ & 0.925 $\pm 0.01$ & 0.978 $\pm 0.01$\\
        \hline
        \multirow{4}{4em}{WNS} & ImgNet & 0.936 $\pm 0.02$ & 0.926 $\pm 0.02$ & 0.984 $\pm 0.004$\\
        & CTP & 0.989 $\pm 0.01$ & 0.989 $\pm 0.01$ & 0.998 $\pm 0.001$ \\
        & TC-Swin & 0.967 $\pm 0.01$ & 0.966 $\pm 0.01$ & 0.993 $\pm 0.01$ \\
        & TC-Conv & 0.973 $\pm 0.01$ & 0.973 $\pm 0.01$ & 0.995 $\pm 0.01$\\
        \hline
        \hline
         \multirow{4}{4em}{Average} & ImgNet & 0.859 $\pm 0.09$ & 0.852 $\pm 0.1$ & 0.962 $\pm 0.02$\\
        & CTP & 0.944 $\pm 0.06$ & 0.943 $\pm 0.01$ & $\mathbf{0.992} \pm 0.01$ \\
        & TC-Swin & $\mathbf{0.953} \pm 0.02$ & $\mathbf{0.953} \pm 0.02$ & 0.989 $\pm 0.01$\\
        & TC-Conv & 0.941 $\pm 0.02$ & 0.940 $\pm 0.02$ & 0.984 $\pm 0.01$\\
    \end{tabular}
    \label{tab:semicol}
\end{table*} 
\autoref{tab:semicol} presents the results per trained center on the SemiCOL challenge dataset. The 'clinic' column indicates the training center. The averaged results are shown at the bottom of the table for orientation. On average the TC-Swin encoder showed higher mean accuracy (0.953) compared to the TC-Conv (0.941) and CTP (0.944) encoders, as well as higher F1-scores. In terms of AUC, all three encoders show approximately the same performance on average.

When looking at the performance on individual clinics, the transfer from UBERN to other clinics seems to be the most challenging. ImageNet (0.734) and CTP (0.860) both showed a performance about 10 percentage points lower than the performance on other centers. However, both TC encoders were able to extract features that allowed the transfer of the learned classification to other centers (TC-Swin: 0.943, TC-Conv: 0.943). The performance when trained on data from the UBERN clinic is similar to that when trained on data from other clinics.

Looking at performance on other clinics, there is a slight discrepancy in mean accuracy and AUC score between CTP and TC. CTP consistently achieves slightly higher scores of approximately one percentage point over the TC-Swin encoder.

\subsection{Lung Cancer}
\begin{table*}[t]
    \caption{Lung: Tissue-specific model performance when applied to the TCGA-NSCLC dataset. The model consist of organ-specific MIL head and frozen encoders: ImageNet, CTP, TC-Swin, TC-Conv. The table shows mean weighted accuracy, F1 score, AUC, and the respective standard deviations over 5 different runs. Each model was trained in a 5-fold cross-validation and in a cross-center setting using the 3 main contributing centers for training and the remaining as hold out test set.}
    \centering
    \begin{tabular}{c|c|c|c|c}
        Evaluation & Model & Acc & F1 & AUC\\
        \hline
        \multirow{2}{4em}{External} &
        \cite{shao_transmil_2021}& 0.884 & - & 0.961 \\
        & \cite{wang_transformer-based_2022} & 0.912 & - & 0.973 \\
        \hline
        \multirow{4}{4em}{5-fold} & ImgNet & 0.838 $\pm 0.02$ & 0.834 $\pm 0.02$ & 0.926 $\pm 0.02$\\
        & CTP & $\mathbf{0.902} \pm 0.02$ & 0.899 $\pm 0.02$ & $\mathbf{0.965} \pm 0.01$ \\
        & TC-Swin & 0.878 $\pm 0.02$ & 0.875 $\pm 0.02$ & 0.949 $\pm 0.01$\\
        & TC-Conv & 0.852 $\pm 0.01$ & 0.847 $\pm 0.01$ & 0.925 $\pm 0.01$\\
        \hline
        \multirow{4}{4em}{cross-center} & ImgNet & 0.733 $\pm 0.01$ & 0.721 $\pm 0.01$ & 0.817 $\pm 0.01$\\
        & CTP & $\mathbf{0.854} \pm 0.01$ & 0.850 $\pm 0.01$ & $\mathbf{0.924} \pm 0.01$ \\
        & TC-Swin & 0.785 $\pm 0.01$ & 0.778 $\pm 0.01$ & 0.863 $\pm 0.01$\\
        & TC-Conv & 0.763 $\pm 0.01$ & 0.759 $\pm 0.01$ & 0.832 $\pm 0.01$\\
    \end{tabular}
    \label{tab:lung}
\end{table*}
\autoref{tab:lung} shows the results of the 5-fold cross-validation and the cross-center evaluation on the TCGA-NSCLC dataset. Despite the absence of lung cancer images in the pre-training, both TC encoders outperform the ImageNet weights in the cross-center and 5-fold cross-validation tasks. The CTP encoder achieves the highest mean accuracy of 0.902, which is about 2 percentage points better than the TC-Swin of 0.878. The CTP encoder was pretrained using the TCGA-NSCLC dataset and its performance achieved when evaluating the CTP in our pipeline is similar to the external baseline reported by \cite{wang_transformer-based_2022} (0.902 vs. 0.912). The TC-Swin encoder shows comparable performance earlier results reported by \cite{shao_transmil_2021}, with an accuracy of 0.878 vs. 0.884 and an AUC of 0.949 vs. 0.961.  

As expected, both models TC-models show weaker performance when evaluated across centers, compared to mixed-center 5-fold cross-validation. Both TC encoders outperform the ImageNet-based encoder. The CTP encoder was pre-trained to extract features from this dataset and outperforms all other models in the cross-center evaluation.

As an additional experiment, the UNI \cite{chen_towards_2024} encoder was compared to the other models using the ABMIL aggregation head and showed similar results to the other encoders. The detailed results can be found in \ref{sec:abmil}.

\section{Discussion}
This paper demonstrates that MTL is an effective method for training foundation models on supervised signals in CPath. We tested the MTL-trained Tissue Concepts encoders on breast, prostate, colorectal, and lung cancers and found that the performance of the encoders is comparable to a self-supervised model. This comparable performance was achieved with only 6\% of the data and resources: training TC required only 912,000 patches, compared to 15 million patches used in \cite{wang_transformer-based_2022}'s self-supervised approach. Less data and shorter training enable faster development and research cycles.

Further development and scaling of the foundation models in CPath will inevitably contribute to extended CO\textsubscript{2} emissions. The TC encoders were trained for 160 hours on a single Nvidia RTX A5000 in Europe. This corresponds to an estimated 18.91 kg of CO\textsubscript{2} emissions \cite{lannelongue_green_2021}. Most self-supervised approaches exceed these emissions by orders of magnitude due to larger training datasets, longer convergence times, and the use of multiple GPUs. Training such a model for the same amount of time on 48 NVIDIA V100s, a common number of GPUs for training large foundation models, results in 2004 kg of CO\textsubscript{2} during training. 
Overall, depending on the GPU, one training of an MTL-based model produces only 0.9\% to 2.25\% of the CO\textsubscript{2} emissions of an SSL-based training while providing similar performance.

Even though, performance generally is comparable, there are two differences between TC and CTP in organ specific performance. One is the difference on the prostate dataset. While a small amount of patches of this dataset were present during the pre-training of original TC encoder, the vast amount of patches used during testing were unseen. Overall, a lot of prostate patches were used during the pre-training of TC, as mentioned in section \ref{sec:DS}. This large proportion of prostate tissue during pre-training may have led to the better performance on this organ  and it  likely prevented a larger reduction of the encoder's performance when omitting the small amount of PANDA data during pre-training.
This raises the question if organ-specific fine-tuning of TC encoders can lead to better performance on specific organs when considering cross-center evaluation and testing on unseen datasets. This question is further justified by the differences in organ specific performance between foundation models and external baselines.

Another clear difference in performance between CTransPath and Tissue Concepts was observed for lung cancer in the 5-fold cross-validation and cross-center validation. This difference most likely arises due to the out-of-domain nature of lung cancer tissue in TC pre-training. However, training on pathology images instead of real-world images improved performance when comparing Tissue Concepts weights to ImageNet weights. Scaling the Tissue Concepts pre-training with more data should be considered to overcome the current limitations. Importantly, balancing different tasks, tissue types, and magnifications during pre-training is important to effectively scale and apply multi-task learning in the CPath domain.

In the present study, only frozen encoders were considered.
While fine-tuning these foundation models can lead to better adapting to specific tasks, more research is needed on the balance between data specificity and variation, and the amount of data is required. In \cite{schafer_overcoming_2023} we found that a fine-tuned foundation model encoder outperformed a fine-tuned ImageNet-base model on all tasks in terms of data efficiency and F1-score. Similar results can be expected for TC encoders and CPath specific problems, but this remains an open question to be investigated further. Additionally, this study focuses on solving MIL problems which occur frequently in diagnostic or prognostic tasks. While MIL tasks are clinically motivated, the individual performance of the segmentation, and detection branches in the multi-task trained model need to be considered in future evaluations.

Overall, the results show that the more domain-specific pre-training of Tissue Concepts encoders is advantageous for solving domain-specific tasks compared to the more general, multi-domain pre-trained encoders. A question that remains to be answered is the performance difference between foundation models and models trained on one specific organ. This will be part of future research.

\section{Conclusion}
In this paper, we propose to train a foundation model for CPath on supervised signals using multi-task learning to reduce the need for large corpora of data, computation time, and resources during training. The proposed method shows comparable results to a model trained using self-supervision while relying on a fraction of the training patches.

In addition, we found that, although trained on large amounts of data, existing models still exhibit a loss of performance when applied across centers. Better cross-center generalization is crucial to facilitate broader clinical application and needs to be further addressed in future research.

In order to effectively scale multi-task learning of foundation models, questions regarding the balance of tasks and tissue types need to be answered. Organ-specific fine-tuning might create robust and high-performing encoders for specific problems but needs to be further explored.

\section{Data availability}
The trained models, scripts, and notebooks to reproduce the results will be made publicly available upon publication.\\
The pre-training framework to build the pipeline will be available as a python package.
The links to all datasets are mentioned in the appendix with a corresponding description.

\section{Declaration of competing interest}
The authors declare that they have no known competing financial or personal relationships or interests that could have appeared to influence the work reported in this paper.

\section{Acknowledgments}
This research was in part funded by the German ministry of education and research (BMBF) through the projects SynDICAD (01IS21067) and PROSurvival (01KD2213). The authors are responsible for the content of this publication.

The results show in this paper are in part based upon data generated by the TCGA Research Network: \url{https://www.cancer.gov/tcga}.
%%Harvard

% \bibliographystyle{model2-names.bst}
% \biboptions{authoryear}
% \bibliography{references} 
\printbibliography

\newpage
\appendix
\section{Training Datasets}
\label{app:DS}

\textbf{NCT-CRC-HE 100k}\\
The NCT-CRC-HE 100k dataset consists of 100.000 non-overlapping patches from 86 H\&E stained human colorectal cancer and normal WSIs. The patches are of size $224 \times 224$ and have a resolution of 0.5 microns per pixel (MPP). Nine different tissue classes were present in this classification dataset. The dataset is available under \url{https://zenodo.org/records/1214456}.

\textbf{Conic}\\
This dataset is based in the Lizard Dataset \cite{graham_lizard_2021}. It contains 4981 H\&E stained, non-overlapping patches of colon tissue at $256 \times 256$ pixels. The segmentation masks were generated using the Hover-Net The dataset is available under \url{https://conic-challenge.grand-challenge.org/}.

\textbf{CRAG}\\
The CRAG dataset consists of 213 H\&E stained patches from colon tissue of 38 WSIs. The patches are of size $1512 \times 1512$ at a resolution of 0.5 MPP. During preprocessing the patches and masks were further cropped into non overlapping $224 \times 224$ tiles with 1MPP resolution. The original 173/40 patch train/val split was maintained. The dataset is available under \url{https://warwick.ac.uk/fac/sci/dcs/research/tia/data/mildnet} with corresponding login.

\textbf{SemiCol}\\
The training data from the SemiCOL challenge is divided into two parts. One contains manual segmentation masks, which were used in the pre-training of the encoder. The other part contains weakly labeled data, which was described in \ref{sec:DS}. The 20 sparsly annotated slides were obtained from the university hospital in Cologne (Hamamatsu S360) and from the university hospital LMU in Munich (Leica GT450). Both sides scanned at 0.5 MPP and provided 10 slides each. Patches with annotations were extracted from the provided images and scaled to 1 MPP and $224 \times 224$ pixels.
More details about the data can be found here \url{https://www.semicol.org/data/}.

\textbf{Arvaniti}\\
The Arvaniti TMA dataset consists of prostate tissue microarrays, which were scanned at 0.23 MPP resolution at the University Hospital Zurich using a Hamamatsu NanoZoomer-XR Digital slide scanner. Patches of  $1024 \times 1024$ were cut out from the foreground and scaled to $224 \times 224$. The major class of the mask was selected to serve as patch label.
The dataset is available under \url{https://dataverse.harvard.edu/dataset.xhtml?persistentId=doi:10.7910/DVN/OCYCMP}.

\textbf{Peso}\\
The PESO dataset consists of 102 WSIs, which were scanned at 20x around 0.48 MPP with a 3DHistech Pannoramic Flash II 250 scanner. From the H\&E stained prostate slides were, patches were extracted and scaled to 1 MPP of size $224 \ times 224$. The corresponding masks were used as targets. 
The dataset is available under \url{https://zenodo.org/records/1485967}.

\textbf{Schoemig-Markiefka}\\
This prostate dataset contains 6 sub-datasets, each containing 120.000 patches. All of the patches were scanned at approximately 0.25 MPP with at least 4 different scanners. Each sub-dataset contains 50.000 patches with tumor tissue, 50.000 patches with non-neoplastic glandular prostate tissue and 20.000 patches with non-glandular tissue. In our training the patches were scaled accordingly. 5 of the sub-datasets were used for pre-training, while 1 was used for validation
The dataset is available under: \url{http://zenodo.org}, Deposits: 4789576 (Dataset 1–4) and 4904569 (Datasets 5–6).

\textbf{Panda}\\
This prostate dataset was already described in \ref{sec:DS}. From the slides, patches of $224 \times 224$ were extracted. The labels for the patches were generated from the corresponding masks.
The dataset is available under \url{https://www.kaggle.com/c/prostate-cancer-grade-assessment/data}.

\textbf{TUH}\\
The Temple University digital pathology corpus consists of over 3505 annotated images of breast tissue. The labels range from artifact and background annotations to specific breast cancer annotations like invasive ductal carcinoma. All WSIs of the 296 patients are scanned at 0.5 MPP with corresponding annotations. From the 3505 slides 136 exhibited quality annotations that were used in the training. Fixed 20 patches per WSI were sampled. 
Details about the download can be found here \url{https://isip.piconepress.com/projects/nsf_dpath/html/downloads.shtml}.

\textbf{Tiger}\\
This breast cancer dataset contains H\&E stained patches from tumor infiltrating lymphocytes. The patches were extracted from 151 TCGA-BRCA slides at f0.25 MPP. Manual annotations for 7 different classes were used as masks. Patches and masks were scaled to $224 \times 224$ pixels.
The dataset is available under \url{https://tiger.grand-challenge.org/Data/}.

\textbf{BCSS}\\
The BCSS dataset was contained as part of the TIGER challenge. Annotations provided through the challenge were used as segmentations masks. These annotations were derived from the original BCSS dataset. All images were scaled same as patches from the Tiger dataset mentioned above. 124 slides from BCSS and NuCLS were annotated, where some of the annotations were grouped.
The dataset is available under \url{https://tiger.grand-challenge.org/Data/}.

\textbf{BreakHis}\\
The breakhis dataset contains H\&E stained patches from 82 patients at different magnifications. The patches are of size $700 \ times 460$ and were scaled accordingly to 1 MPP resolution. The classes for this dataset were derived from the original dataset. A two-class benign/malignant split is possible, however, the more fine-grained 8 class dataset was used.
This dataset is available under \url{https://web.inf.ufpr.br/vri/databases/breast-cancer-histopathological-database-breakhis/}.

\textbf{MiDoG}\\
The MIDOG 2022 challenge dataset contains 405 tumor cases across six different tumor types. During training the 44 lung cancer cases were excluded. The cases were patchified and scaled to $224 \ times 224$ pixels with approximately 1 MPP.
The dataset is available under \url{https://zenodo.org/records/6547151}.

\textbf{HubMap}\\
The HubMap dataset contains 351 cases from different organs. Patches were extracted from the foreground containing annotation masks and scaled to 1 MPP of $224 \times 224$ pixels. Only large intestine was used during training.
The dataset can be found under \url{https://www.kaggle.com/datasets/dingyan/hubmap-data}.

\section{MIL Head Architecture}
\label{sec:mil-head}
To aggregate the latent WSIs, as described in section \ref{sec:wsi_clf}, a small CNN with global max pooling, based on \cite{tellez_neural_2021} was used. The initial layer utilized a 1x1 convolution, reducing the latent WSIs' 768 channel to 16. Following this, 3x3 convolutions with a stride of 2, and padding of 1 doubling in depth each time were applied. Instance normalization, leaky ReLU activation, and 10\% 2D dropout were applied in the convolutional layers. The lowest feature pyramid of 64 channels was globally max pooled, followed by one hidden linear layer and a classification layer. This resulted in 19250 total trainable parameters. The convolutional layers were initialized by drawing from a Kaiming normal distribution with a fixed seed. 

\subsection{ABMIL Results and comparison to the UNI encoder}
\label{sec:abmil}
A different aggregation approach was proposed by \cite{ilse2018attention} under the term "Attention-based Deep Multiple Instance Learning”. For comparison, the following tables present the results obtained by using an attention based multiple instance learning aggregation head instead of a maximum pooling aggregation head. The results were collected over three runs and averaged. Additionally, the experiments were repeated using the UNI \cite{chen2024towards} foundation model.

Overall, we observed similar results between the different encoders and aggregation techniques across all organs. 
%%% BREAST
\begin{table*}[t]
    \caption{Breast: Tissue-specific model performance when applied to the BRACS challenge data. The model consist of organ-specific head and frozen encoders: ImageNet, CTP, TC-Swin, TC-Conv, UNI. The table shows mean F1 score, AUC, Balanced Accuracy, and standard deviation over 3 different runs for different problem formulations. All results were obtained using an ABMIL aggregation head.}
    \label{tbl:appendix_breast}
    \centering
    \begin{tabular}{c|c|c|c|c}
        Problem & Model &  F1 & AUC & b ACC \\
        \hline
         \multirow{5}{4em}{2-class} & \cite{pati_hierarchical_2022} & \textit{0.96} & -  & -\\
         \cline{2-5}
         & ImageNet & $0.888 \pm 0.04$ & $0.968  \pm 0.002$  & $0.865 \pm 0.05$\\
         & CTP & $0.884 \pm 0.05$  & $0.990 \pm 0.001$ & $ 0.853 \pm 0.06$\\
         & TC-swin & $0.874 \pm 0.02$ & $0.983 \pm 0.02$ & $0.833 \pm 0.05$\\
         & TC-conv & $0.893 \pm 0.09$ & $0.990 \pm 0.001$ & $ 0.870 \pm 0.001$\\
         & UNI & $0.930 \pm 0.03$ & $0.992 \pm 0.006$ & $0.901 \pm 0.04$\\
        \hline
        \multirow{5}{4em}{4-class} & \cite{pati_hierarchical_2022} & \textit{0.84} & -  & - \\
         \cline{2-5}
         & ImageNet & $0.283 \pm 0.01$ & $0.677 \pm 0.05$ & $0.357 \pm 0.02$  \\
         & CTP & $0.329 \pm 0.01$ & $0.759 \pm 0.01$ & $0.414 \pm 0.01$ \\
         & TC-swin & $0.313 \pm 0.04$ & $0.735 \pm 0.05$ & $0.381 \pm 0.04$\\
         & TC-conv & $0.276 \pm 0.01$ & $0.699 \pm 0.01$ & $0.347 \pm 0.02$\\
         & UNI & $0.287 \pm 0.02$ & $0.711 \pm 0.06$ & $ 0.357 \pm 0.02$\\
         \hline
         \multirow{5}{4em}{7-class} & \cite{pati_hierarchical_2022} & \textit{0.69} & -  & -\\
         \cline{2-5}
         & ImageNet & $0.189 \pm 0.01$ & $0.667 \pm 0.001$ & $0.275 \pm 0.01$\\
         & CTP & $0.186 \pm 0.007$ & $0.698 \pm 0.001$ & $0.278 \pm 0.01$ \\
         & TC-swin & $0.180 \pm 0.01$ & $0.685 \pm 0.02$ & $0.275 \pm 0.009$\\
         & TC-conv & $0.115 \pm 0.06$ & $0.580 \pm 0.06$ & $0.194 \pm 0.08$\\
         & UNI & $0.183 \pm 0.03$ & $0.674 \pm 0.02$ & $0.269 \pm 0.01$\\
    \end{tabular}
\end{table*}

%%% PROSTATE
\begin{table*}
    \caption{Prostate: Tissue-specific model performance when applied to the PANDA challenge data. The model consist of organ-specific head and frozen encoders: ImageNet, CTP, TC-Swin, TC-Conv. The table shows mean weighted Accuracy, F1 score, AUC, balanced Accuracy, and standard deviations over 3 different runs for 5-fold cross-validation and cross-center transfer performance. All results were obtained using an ABMIL aggregation head.}
    \centering
    \begin{tabular}{c|c|c|c|c|c|c}
         Clinic & Model & Acc & F1 & AUC & Balanced Acc\\
         \hline
         external & \cite{mahdi_behzadi_weakly-supervised_2022}& 0.95 & 0.93 & - & -\\
         \hline
         \multirow{4}{4em}{5 fold} 
         & ImgNet & $0.823 \pm 0.07$ & $0.793 \pm 0.06$ & $0.900 \pm 0.03$ & $0.819 \pm 0.06$\\
        & CTP & $0.865 \pm 0.03$ & $0.834 \pm 0.05$ & $0.924 \pm 0.04$ & $0.847 \pm 0.06$\\
        & TC-Swin & $0.923 \pm 0.01$ & $0.905 \pm 0.02$ & $0.970 \pm 0.01$ & $0.913 \pm 0.03$  \\
        & TC-Conv &  $0.923 \pm 0.02$ & $0.904 \pm 0.02$ & $0.969 0.02$ & $0.914 \pm 0.03$\\
        & UNI & $0.944 \pm 0.01$ & $0.931 \pm 0.01$ & $0.982 \pm 0.01$ & $0.939 \pm 0.02$\\
        \hline
         \multirow{4}{4em}{Karolinska $\rightarrow$ Radboud} 
         & ImgNet & $0.595 \pm 0.07$ & $0.576 \pm 0.06$ & $0.662 \pm 0.07$& $0.618 \pm 0.07$\\
        & CTP & $0.819 \pm 0.02$ & $0.605 \pm 0.02$ & $0.825 \pm 0.01$ & $0.627 \pm 0.02$\\
        & TC-Swin & $0.902 \pm 0.01$ & $0.857 \pm 0.01$ & $0.948 \pm 0.01$ & $0.904 \pm 0.01$  \\
        & TC-Conv & $0.726 \pm 0.01$ & $0.684 \pm 0.01$ &  $0.919\pm 0.01$ & $0.816 \pm 0.01$ & \\
        & UNI & $0.748 \pm 0.01$ & $0.702 \pm 0.01$ & $0.893 \pm 0.02$ & $0.825 \pm 0.01$ \\ 
        \hline
        \multirow{4}{4em}{Radboud $\rightarrow$ Karolinska} 
        & ImgNet & $0.572 \pm 0.07$ & $0.572 \pm 0.07$ & $0.668 \pm 0.04$ & $0.626 \pm 0.07$\\
        & CTP & $0.676 \pm 0.03$ & $0.583 \pm 0.07$ & $0.729 \pm 0.01$  & $0.595 \pm 0.03$  \\
        & TC-Swin & $0.896 \pm 0.01$ & $0.891\pm 0.01$ & $0.969 \pm 0.04$& $0.915 \pm 0.01$ \\
        & TC-Conv & $0.654 \pm 0.05$ & $0.676 \pm 0.06$ & $0.898 \pm 0.01$ & $0.902 \pm 0.01$ \\
        & UNI & $0.662 \pm 0.01$ & $0.661 \pm 0.01$	& $0.871 \pm 0.02$ & $0.738 \pm 0.02$\\
    \end{tabular}
\end{table*}

%%% COLORECTAL
\begin{table*}[t]
    \centering
    \caption{Colorectal: Tissue-specific model performance when applied to the SemiCOL challenge data. The model consist of organ-specific MIL head and frozen encoders: ImageNet, CTP, TC-Swin, TC-Conv, UNI. The table shows mean accuracy, F1 score, AUC, Balanced Accuracy, and the respective standard deviations over 3 different runs. Each model was trained on images from one site and evaluated on all images from the remaining sites. All results were obtained using an ABMIL aggregation head.}
    \begin{tabular}{c|c|c|c|c|c}
        Clinic & Model & Acc & F1 & Auc & Balanced Acc\\
        \hline
        \multirow{4}{4em}{LMU} 
        & ImgNet & $0.678 \pm 0.12$ & $0.646 \pm 0.16$ & $0.741 \pm 0.16$ & $0.678 \pm 0.12$\\
        & CTP & $0.925\pm 0.01$ &$0.925 \pm 0.01$ & $0.985 \pm 0.005$  & $0.925 \pm 0.01$\\
        & TC-Swin & $0.873 \pm 0.06$ & $0.871\pm 0.07$ & $0.934 \pm 0.07$  & $0.873 \pm 0.06$\\
        & TC-Conv & $0.913 \pm 0.06$ & $0.912 \pm 0.06$ & $0.981 \pm 0.02$  & $0.913 \pm 0.06$\\
        & UNI & $0.807 \pm 0.03$ & $0.802 \pm 0.04$ & $0.939 \pm 0.02$  & $0.809 \pm 0.03$\\
        \hline
        \multirow{4}{4em}{UBERN}
        & ImgNet & $0.793 \pm 0.18$ & $0.793 \pm 0.18$ & $0.838 \pm 0.20$ & $0.793 \pm 0.18$\\
        & CTP & $0.820 \pm 0.03$ & $0.819 \pm 0.04$ & $0.914 \pm 0.02$ & $0.820 \pm 0.03$\\
        & TC-Swin & $0.915 \pm 0.01$ & $0.915 \pm 0.01$ & $0.973 \pm 0.01$ & $0.915 \pm 0.01$\\
        & TC-Conv &$0.947 \pm 0.01$ & $0.947 \pm 0.01$ & $0.984 \pm 0.01$ & $0.947 \pm 0.01$\\
         & UNI & $0.905 \pm 0.04$ & $0.904 \pm 0.04$ & $0.956 \pm 0.03 $& $0.905 \pm 0.04$\\
        \hline
        \multirow{4}{4em}{UKK}
        & ImgNet & $0.865 \pm 0.064$& $0.864\pm 0.064$ & $0.927 \pm 0.067$ & $0.866 \pm 0.064$ \\
        & CTP &  $0.901 \pm 0.104$ & $0.897 \pm 0.109$ & $0.979 \pm  0.024$  & $0.901 \pm 0.103$ \\
        & TC-Swin & $0.809 \pm 0.099$ &	$0.799 \pm 0.112$ &	$0.965 \pm 0.027$ &	$0.809\pm 0.098$ \\
        & TC-Conv & $0.892\pm 0.050$ &$0.892 \pm 0.050$& $0.953 \pm 0.050$& $0.892 \pm 0.050$\\
         & UNI & $0.835 \pm 0.082$ & $0.833 \pm 0.082 $& $0.935 \pm 0.68$ &$0.834 \pm 0.081$\\
        \hline
        \multirow{4}{4em}{WNS}
         & ImgNet & $0.801 \pm  0.22$& $0.801 \pm 0.22$& $0.857 \pm 0.22$ & $0.801 \pm 0.22$ \\
        & CTP & $0.962 \pm 0.01$ &$ 0.962 \pm 0.01$ & $0.992 \pm 0.02$ & $0.962 \pm 0.01$ \\
        & TC-Swin &  $0.949 \pm 0.02$ & $0.949 \pm 0.02$ & $0.993 \pm 0.02$ & $0.949 \pm 0.02$\\
        & TC-Conv &  $0.960 \pm 0.01$ & $0.960 \pm 0.01$ & $0.994 \pm 0.01$ &$ 0.960 \pm 0.01$\\
         & UNI &  $0.964 \pm 0.02$ & $0.964 \pm 0.02$ & $0.993 \pm 0.01 $& $0.964 \pm 0.02$ \\
        \hline
        \hline
         \multirow{4}{4em}{Average} 
        & ImgNet & $0.783 \pm 0.16$ & $0.774 \pm 0.16$ & $0.839 \pm 0.15$ & $0.783 \pm 0.14$ \\
        & CTP & $0.902\pm 0.07$ & $0.901 \pm 0.07$ & $0.968 \pm 0.04$ & $0.902 \pm 0.07$\\
        & TC-Swin & $0.887 \pm 0.09$& $0.884 \pm 0.09$ & $0.966 \pm 0.07$ & $0.887 \pm 0.09$ \\
        & TC-Conv & $0.900 \pm 0.08$ & $0.899 \pm 0.08$ & $0.955 \pm 0.04$ & $0.900 \pm 0.08$\\
         & UNI & $0.88 \pm 0.08$ & $0.880 \pm 0.08$ & $0.958 \pm 0.04$& $0.882 \pm 0.08$ \\
    \end{tabular}
\end{table*} 

%%% LUNG
\begin{table*}[t]
    \caption{Lung: Tissue-specific model performance when applied to the TCGA-NSCLC dataset. The model consist of organ-specific MIL head and frozen encoders: ImageNet, CTP, TC-Swin, TC-Conv, UNI. The table shows mean accuracy, F1 score, AUC, Balanced Accuracy, and the respective standard deviations over 3 different runs. Each model was trained in a 5-fold cross-validation and in a cross-center setting using the 3 main contributing centers for training and the remaining as hold out test set. All results were obtained using an ABMIL aggregation head.}
    \centering
    \begin{tabular}{c|c|c|c|c|c}
        Evaluation & Model & Acc & F1 & AUC & Balanced Acc\\
        \hline
        \multirow{2}{4em}{External} &
        \cite{shao_transmil_2021}& 0.884 & - & 0.961 & - \\
        & \cite{wang_transformer-based_2022} & 0.912 & - & 0.973 & -\\
        \hline
        \multirow{4}{4em}{5-fold} 
        &  ImgNet & $0.803 \pm 0.0001$& $0.807 \pm 0.0001$ & $0.0890 \pm 0.0001$ & $0.801 \pm 0.0001$\\
        & CTP & $0.889 \pm 0.001$ & $0.888 \pm 0.001$ & $0.954 \pm 0.001$ & $0.887 \pm 0.001$ \\
        & TC-Swin & $0.843 \pm 0.001$ & $0.830 \pm 0.001$ & $0.947 \pm 0.001$ & $0.840 \pm 0.001$\\
        & TC-Conv & $0.865 \pm 0.001$ & $0.864 \pm 0.001$ & $0.933 \pm 0.001$ & $0.865 \pm 0.001$\\
        & UNI & $0.924 \pm 0.001$ & $0.923 \pm 0.001$ & $0.970 \pm 0.001$ & $0.920 \pm 0.001$ \\
        \hline
        \multirow{4}{4em}{cross-center}
        & ImgNet & $0.723 \pm 0.0001$ & $0.720 \pm 0.0001$ & $0.777 \pm 0.0001$ & $0.720 \pm 0.0001$\\
        & CTP & $0.804 \pm 0.0001$ & $0.802 \pm 0.0001$ & $0.891 \pm 0.0001$ & $0.801 \pm 0.0001$ \\
        & TC-Swin & $0.786 \pm 0.0001$ & $0.772 \pm 0.0001$ & $0.859 \pm 0.0001$ & $0.771 \pm 0.0001$\\
        & TC-Conv & $0.720 \pm 0.0001$& $0.718 \pm 0.0001$& $0.784 \pm 0.0001$ & $0.719 \pm 0.0001$\\
        & UNI & $0.857 \pm 0.0001$ & $0.855 \pm 0.0001$ & $0.917 \pm 0.0001$ & $0.852 \pm 0.0001$ \\
    \end{tabular}
\end{table*}

\end{document}